\documentclass[twocolumn,amsmath,amssymb]{revtex4-1}
\usepackage{txfonts}

\usepackage{bm}
\usepackage[pdftex]{graphicx}
\usepackage{braket}

\usepackage{mathdots}

\newcommand  {\eqn}[1]{(\ref{eqn:#1})}
\renewcommand{\(}     {\left(}
\renewcommand{\)}     {\right)}
\renewcommand{\[}     {\left[}
\renewcommand{\]}     {\right]}
\renewcommand{\_}[1]  {_\textrm{#1}}

\begin{document}

\title{
Helicity-Protected Domain-Wall Magnetoresistance in Ferromagnetic Weyl Semimetal
}
\author{Koji Kobayashi}
\author{Yuya Ominato}
\author{Kentaro Nomura}
\affiliation{Institute for Materials Research, Tohoku University, Sendai Aoba-ku 980-8577, Japan}

\begin{abstract}
 The magnetotransport properties of disordered ferromagnetic Weyl semimetals are investigated numerically.
 We found an extraordinarily stable and huge magnetoresistance effect in domain walls of Weyl semimetals.
 This effect originates from the helicity mismatch of Weyl fermions and is a specific property of Weyl semimetals.
 Although conventional magnetoresistance effects are strongly suppressed in domain walls where local magnetization varies gradually, 
the helicity-protected magnetoresistance in Weyl semimetals
maintains almost $100\%$ of the magnetoresistance ratio for any kind of thick domain walls, even in the presence of disorder.
 The contribution of surface Fermi arcs to the magnetoresistance is also discussed.
\end{abstract}

\maketitle


 The magnetoresistance effect has been utilized to read magnetic data in hard disk drives \cite{Parkin1995}.
 Giant magnetoresistance (GMR) occurs in magnetic/nonmagnetic/magnetic trilayer structures 
because 
the spins of the conduction electrons lag behind in their orientation with respect to the local magnetization direction,
which changes abruptly from one magnetic layer to another.
 Materials with high spin-polarization (i.e., half-metals) 
have been searched for with an aim to enhance the magnetoresistance effect \cite{Pickett01}.

\begin{figure}[!htbp]
 \centering
  \includegraphics[width=\linewidth, bb =0 0 1025 1210]{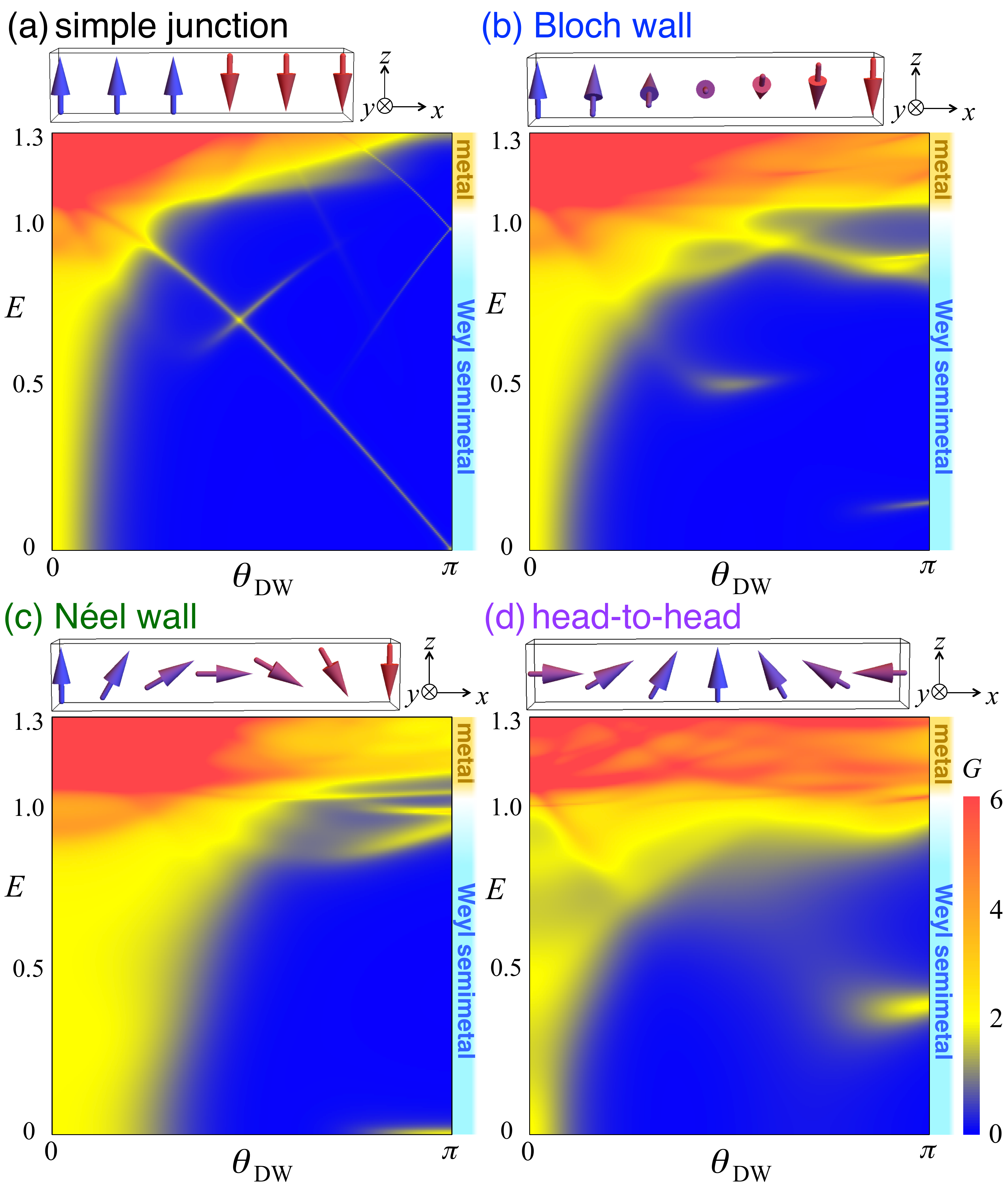}
 \vspace{-7mm}
\caption{
  Conductance maps for the clean bulk of magnetic WSMs ($N=12$) 
  with four types of domain walls: 
  (a) simple junction, 
  (b) Bloch, 
  (c) N{\'e}el, and 
  (d) head-to-head. 
  The vertical axis is the Fermi energy $E$, and the horizontal axis is the relative angle of the magnetization $\theta\_{DW}$ between both ends of the wall 
[see also Fig.~\ref{fig:schematic}(a)].
 In the blue region, the conductance is vanishing and huge magnetoresistance arises.
 The small peak (yellow) structures in the maps originate from the lattice structure 
 and are qualitatively inessential \cite{sup1}.
}
\label{fig:map}
\end{figure}

 Domain-wall magnetoresistance is a similar effect 
that occurs in single magnetic materials with magnetic domain walls \cite{Kent01}.
 Compared to that in trilayer structures, 
the domain-wall magnetoresistance effect is strongly suppressed
as the wall thickness increases
because the spatial change of local magnetization is much more gradual in the presence of domain wall structures.
 However, recent developments in the electronic control of domain walls 
has renewed interest in electronic transport through domain walls 
\cite{MaekawaBook}.
 In this paper, we propose a novel type of magnetoresistance effect
that is not at all suppressed by the thick domain walls (see Fig.~\ref{fig:map})
in ferromagnetic Weyl semimetals (WSMs)
 \cite{Wan11,Burkov11}.
 Moreover, 
this magnetoresistance effect is robust against disorder,
and is considered to originate from the peculiar transport properties of WSMs:
the helicity dependent transport.

 Weyl semimetals form a class of topological materials that
realize the three-dimensional (3D) Weyl fermion systems near the Weyl nodes. 
 That is, the states near a Weyl node $\bm{k}_0$ are described by the effective Hamiltonian 
\begin{align}
 H\_{Weyl}(\bm{k}) = v\, \bm{\sigma} \cdot (\bm{k}-\bm{k}_0)\,.
 \label{eqn:Weyl}
\end{align}
 The WSMs are typically realized by breaking 
either 
the time-reversal or inversion symmetry of Dirac semimetals.
 Although the WSM materials discovered to date (such as TaAs, TaP, and NbP \cite{Huan15,Weng15,Xu15,Souma16}) 
are the inversion-broken type,
candidates for the time-reversal broken type, 
i.e., 
the ferromagnetic WSMs
have been proposed recently: magnetic Heusler compounds \cite{Hirschberger16,Wang16}, compounds with tetragonal structures \cite{Jin17},
and Co$_3$Sn$_2$S$_2$ 
\cite{eLiu17}.
 (Note that antiferromagnetic type WSMs may be realized in Y$_2$Ir$_2$O$_7$ \cite{Wan11}, YbMnBi$_2$ \cite{Borisenko15}, and Mn$_3$Sn \cite{Yang17,Ito17}.)
 The Dirac semimetals arise on the phase boundary between topologically different insulator phases.
 This bulk gapless region is broadened by breaking 
 one of the symmetries 
\cite{Murakami07,Bulmash14,Kurebayashi14}, 
and is transformed into the WSM phase.
 This means that ferromagnetic WSMs may be achieved by magnetically doping the topological insulators, e.g., Bi$_2$Se$_3$.
 Such a doped material is essentially inhomogeneous,
and thus the effect of disorder must be taken into account.


 We employ a simple 3D lattice model for ferromagnetic WSMs
based on the Wilson-Dirac type tight-binding Hamiltonian
\cite{Liu10,Ryu12},
which describes the 3D topological insulators on a cubic lattice, 
and exhibits a Dirac semimetal phase between the topological and ordinary insulating phases.
 By introducing
the exchange coupling term,
we obtain 
the Hamiltonian for magnetic WSMs,
\begin{align}
 H &=  \sum_{\bf r} \sum_{\mu=x,y,z} 
        \[ c^{\dag}_{{\bf r}+{\bf e}_\mu} 
           \(  {\textrm{i}t \over 2}  \alpha_{\mu}
              -{m_2 \over 2} \beta 
           \)
           c_{\bf r}  + \textrm{h.c.}
        \]  \nonumber \\
   & + \sum_{\bf r} c^{\dag}_{\bf r} 
        \[(m_0 +3 m_2) \beta
           +\bm{M}({\bf r}) \cdot \bm{S}
           + V({\bf r}) 1_4
        \] c_{\bf r},
 \label{eqn:H_real}
\end{align}
where ${\bf r}$ is the position of lattice sites and ${\bf e}_\mu$ 
($\mu = x,y,z$)
is the lattice vector in the $\mu$ direction.
 $m_0$ is referred to as ``mass'' and $m_2$ is the Wilson term.
 The length unit is set to the lattice constant.
$\alpha_\mu$ and $\beta$ are an anticommuting set of matrices 
and $\alpha_\mu^2 = \beta^2 = 1_4$.
 We choose the explicit representation of these matrices as
$\bm{\alpha} = \tau_z \otimes \bm{\sigma}$ and $\beta   = \tau_x \otimes 1_2$,
where $\bm \sigma$ and $\bm \tau$ are the Pauli matrices,
which correspond to the real- and pseudo-spin degrees of freedom.
 Therefore, the spin operator is represented as $\bm{S} = 1_2 \otimes \bm{\sigma}$.
 We introduce an on-site random potential $V({\bf r})$, which is uniformly distributed in $[-{W\over 2},{W\over 2}]$.
 The parameters $t = 2$, $m_2 = 1$, $m_0 = 0$ 
 (see Ref.~\onlinecite{note1}%
),
and $|M| \simeq 1$ are set so that a single pair of Weyl nodes appears at $k_0 = \pm{\pi \over 6}$ (see Ref.~\onlinecite{note2}%
).
 We note that the Weyl nodes should be sufficiently separated for the validity of the discussion below.

\begin{figure}[tbp]
 \centering
  \includegraphics[width=0.98\linewidth, bb =0 0 483 503]{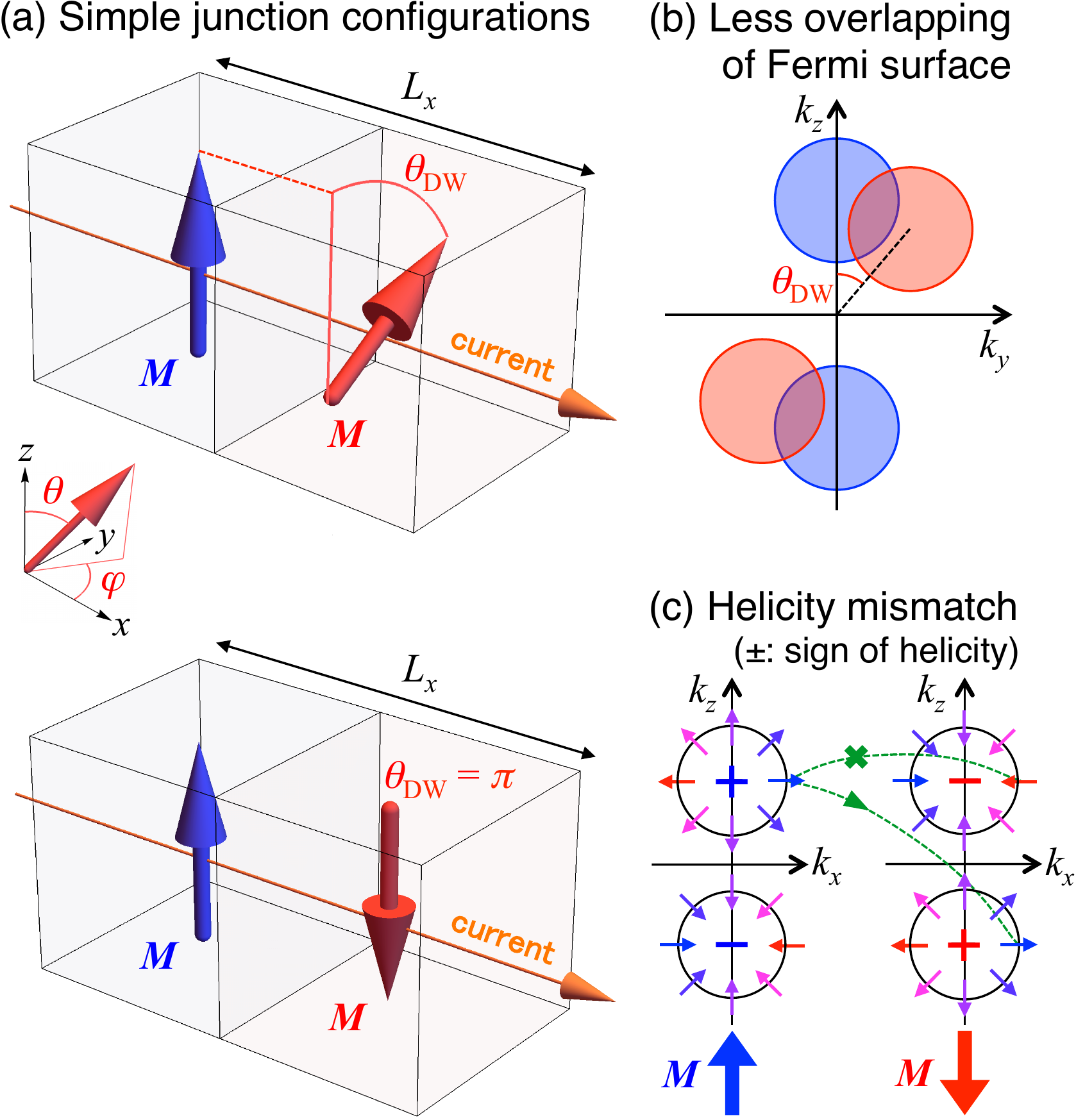}
 \vspace{-3mm}
\caption{
  (a) Schematic figures of simple junction systems.
  $\theta\_{DW}$ denotes the relative angle of the magnetizations in two WSMs.
  Schematic images of (b) less overlapping of projected Fermi surfaces and 
  (c) helicity mismatch of the overlapping nodes. 
  The small colored arrows indicate the electron spin.
  The helicities of the Weyl electrons are opposite for overlapping Fermi surfaces
  and therefore the transmission is suppressed in the absence of inter-node scattering.
}
\label{fig:schematic}
\end{figure}

 We study the transport through 
(a) the junction of two magnetic WSMs with different magnetization directions
and through three types of domain walls: (b) Bloch, (c) N{\'e}el, and (d) head-to-head type
(see also schematic magnetic textures in Fig.~\ref{fig:map}).
 The direction of the current is set to be the $x$ axis.
 The magnetic structure can be implemented by the $x$-position dependent magnetization 
$\bm{M}(x)=  M (\sin\theta \cos\varphi, \sin\theta \sin\varphi , \cos\theta )$
[see Fig.~\ref{fig:schematic}(a)].
 The magnitude of magnetization $M$ is assumed to be uniform.
 Each magnetic structure is achieved with a fixed $\varphi$ and $x$-position dependent rotation angle $\theta$, 
and these are defined for
(a) the simple junction ($+z$$|$$-z$),
\begin{alignat}{2}
 \theta(x) =
 \begin{cases}
    \ 0            &  (0\leq x < \frac{L_x}{2})\\
    \ \theta\_{DW} &  (\frac{L_x}{2} < x \leq L_x)
 \end{cases},
 \ \varphi = \pi/2;
 \label{eqn:junction}
\end{alignat}
for (b) the Bloch wall ($+z \to +y \to -z$),
\begin{align}
 \theta(x) =  {x \over L_x}\theta\_{DW},\ \varphi = {\pi\over2};
 \label{eqn:Bloch}
\end{align}
for (c) the N{\'e}el wall ($+z \to +x \to -z$),
\begin{align}
 \theta(x) =  {x \over L_x}\theta\_{DW},\ \varphi = 0;
 \label{eqn:Neel}
\end{align}
and for (d) the head-to-head wall ($+x \to +z \to -x$),
\begin{align}
 \theta(x) =  -{x \over L_x}\theta\_{DW}+{\pi\over 2},\ \varphi = 0.
 \label{eqn:h2h}
\end{align}
 We consider cubic samples with $N\times N\times N$ sites, 
and
numerically calculate the two-terminal conductance 
between the ideal metallic leads attached to $x=0$ and 
$x=L_x=N-1$
by using the transfer matrix method
\cite{Kobayashi13,Kobayashi15} and the Landauer formula.

 First we focus on the transport in the clean bulk 
by imposing periodic boundary conditions in the $y$ and $z$ directions and 
setting $W=0$.
 In a uniformly magnetized system (i.e., $\theta\_{DW}=0$),
the transport near the Weyl nodes at energy $E=0$ relies only on the Weyl cones.
 Hence, in the range $-{2\pi t \over N}\lesssim E\lesssim {2\pi t \over N}$ 
(${2\pi t \over N} \simeq 1$ in Fig.~\ref{fig:map}), 
there is only a single pair of conducting channels that correspond to the lowest band ($k_y=k_z=0$) of the Weyl cones
and the quantized conductance $G=2$ (in units of $e^2/h$) arises [yellow plateaus on $\theta\_{DW}=0$ in Figs.~\ref{fig:map}(a)--(c)].
 For the highly doped case 
$|E|\gtrsim {t - M}$ ($\simeq 1$ in this paper), 
the bulk metallic bands dominate and the feature of WSMs disappears (red regions in Fig.~\ref{fig:map}).

 In the junction system consisting of two magnetic WSMs with different directions of magnetization,
the conductance decreases
as the relative angle of the magnetizations $\theta\_{DW}$ increases
in 
the low energy region ($E<1$)
 [Fig.~\ref{fig:map}(a)].
 This significant reduction of transport 
 from $\theta\_{DW}=0$ to $\theta\_{DW}={\pi \over 2}$
 can be understood as \textit{less overlapping} of the Fermi surfaces \cite{Ominato17,Nguyen06}.
 That is, by changing the magnetization direction,
the position of the Weyl points and the Fermi surfaces enclosing them shifts from the original position
[Figs.~\ref{fig:schematic}(a) and \ref{fig:schematic}(b)].
 As a result, the overlapping area of the projected Fermi surfaces of the two WSMs (i.e., number of current carrying states) decreases.
 On the other hand, this simple picture cannot explain the conductance behavior around $\theta\_{DW}=\pi$; 
the conductance remains strongly suppressed,
even though the Fermi surfaces are again completely overlapped at $\theta\_{DW}=\pi$
[Figs.~\ref{fig:schematic}(a) and \ref{fig:schematic}(c)].
 This implies that the current is almost perfectly reflected at the interface of the WSM with antiparallel magnetizations due to an unconventional reason:
the \textit{helicity mismatch} of the Weyl electrons.
 The Weyl fermion state characterized by the Hamiltonian Eq.~\eqn{Weyl} is an eigenstate of the helicity operator 
${ \bm{\sigma}\cdot\bm{p} \over |\bm{p}| } = { \bm{\sigma}\cdot(\bm{k}-\bm{k}_0) \over |\bm{k}-\bm{k}_0| }$.
 The sign of the helicity is locked around a Weyl node and opposite to that for the partner
\cite{Weyl29,Balents11}
 [illustrated in Fig.~\ref{fig:schematic}(c)].
 For an antiparallel junction, 
the helicities of the 
Weyl electrons in the overlapping Fermi surfaces are opposite
and current cannot pass through the junction.

 We next investigate how the type of domain walls affects the transport.
 The conductance maps in Fig.~\ref{fig:map} show that the qualitative behavior 
in a system with domain walls (b)--(d)
is the same as that in (a) the simple junction;
the conductance decreases as $\theta\_{DW}$ increases and vanishes at $\theta\_{DW}=\pi$.
 We note that the thickness of the wall does not play an essential role
for (b) Bloch and (c) N{\'e}el walls, where the thin-wall limit corresponds to the (a) simple junction.
 In these domain walls, the combined effect of the helicity mismatch and less overlapping suppresses the transport \cite{sup2},
and the conductance remains vanishing for any wall thickness.
 In contrast, for (d) the head-to-head wall,
the system is conducting
in the thin-wall limit, where the helicity mismatch does not work.
 However, the conductance decays exponentially 
as the thickness of the wall increases
because the less overlapping mechanism works in a thick domain wall.
 As a result, a huge resistance arises in a thick domain wall between antiparallelly magnetized domains, 
irrespective of the details of the magnetic texture.

\begin{figure}[tbp]
 \centering
  \includegraphics[width=1\linewidth, bb =0 0 674 296]{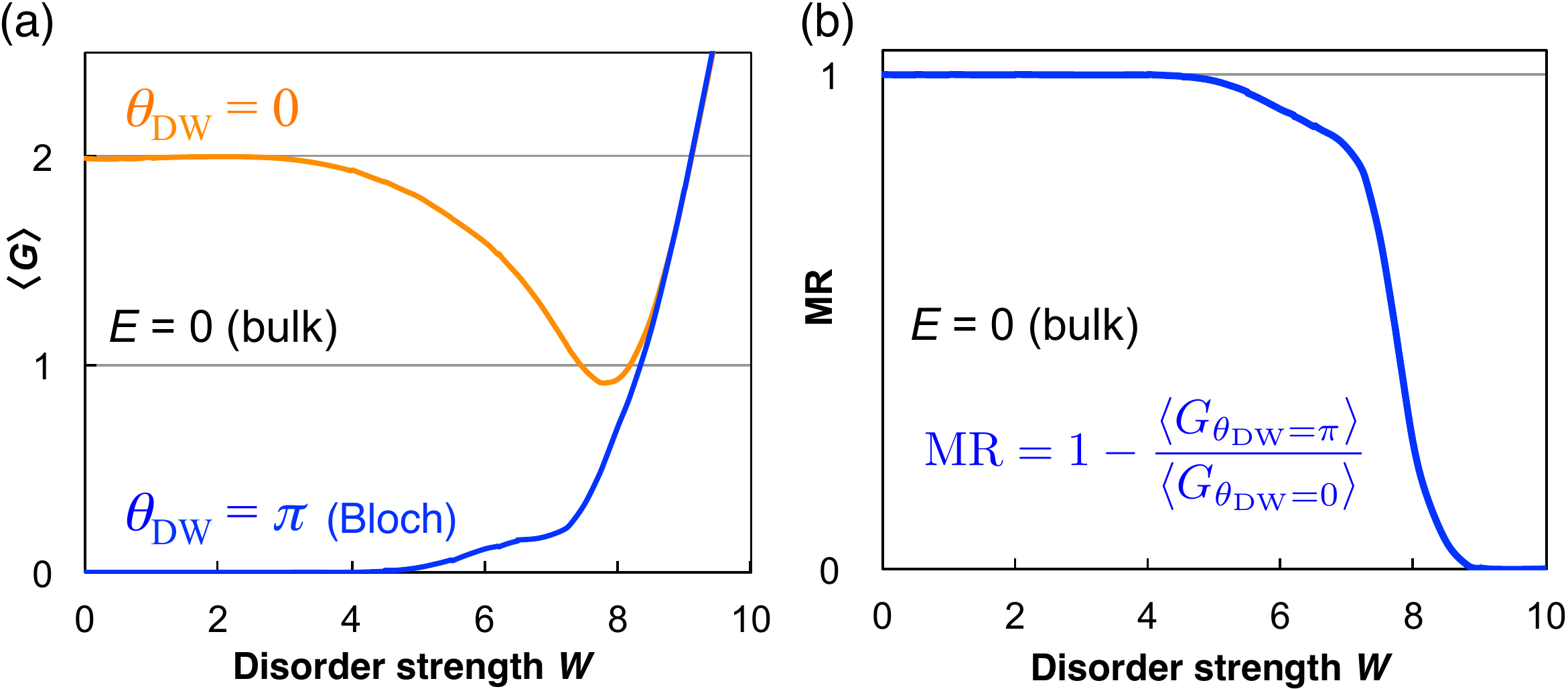}
 \vspace{-5mm}
\caption{
  (a) Averaged conductance $\braket{G}$ as a function of disorder strength $W$, 
 for uniform magnetization ($\theta\_{DW}=0$) and antiparallel ($\theta\_{DW}=\pi$) magnetization with a Bloch wall.
 (b) Magnetoresistance ratio MR as a function of $W$.
 We set $N=24$, and Fermi energy $E=0$.
 Throughout this paper,
 each data point is an average over more than $2000$ disorder realizations and the statistical error is less than $0.01 e^2/h$.
}
\label{fig:G-W}
\end{figure}

 Then we discuss 
the effect of disorder on the transport.
 Figure \ref{fig:G-W}(a) shows the conductance near the Weyl nodes ($E=0$) for uniform magnetization ($\theta\_{DW}=0$) and 
antiparallel configuration with a Bloch wall ($\theta\_{DW}=\pi$).
 The transport in WSMs is robust against disorder, 
so that 
the conductance for uniform magnetization shows a well-quantized plateau, even in the presence of disorder.
 As the disorder strength increases further,
the plateau breaks down 
due to inter-node scattering 
\cite{Kobayashi13}.
 The conductance for the antiparallel configuration
remains suppressed up to a certain strength of disorder ($W\simeq 5$),
and increases
as the disorder strength increases further.
 The difference of the conductance between the parallel and antiparallel cases 
disappears at strong disorder ($W\gtrsim 8$) 
where the system goes into the diffusive metallic phase
from the WSM phase 
\cite{Kobayashi14,Liu16}.
 To characterize the magnitude of the 
magnetoresistance effect,
we introduce the magnetoresistance ratio (MR),
which is defined as
\begin{align}
 \textrm{MR} &= 1-{\braket{G_{\theta\_{DW}=\pi}} \over \braket{G_{\theta\_{DW}=0}} },
 \label{eqn:MR}
\end{align}
where $\braket{\cdots}$ represents an ensemble average.
 Using this quantity, we replotted the data in Fig.~\ref{fig:G-W}(b).
 At weak disorder ($W\lesssim5$), the MR remains at almost $100\%$.
 This shows that 
the magnetoresistance effect is stable even in the presence of weak disorder.
 At strong disorder ($W\gtrsim 8$),
say, in the diffusive metallic phase,
the MR vanishes.
 Although this upper bound of disorder strength is dependent on the Fermi energy $E$,
the qualitative behavior is the same for $|E|\lesssim 1$.

\begin{figure}[tbp]
 \centering
  \includegraphics[width=1\linewidth, bb =0 0 772 491]{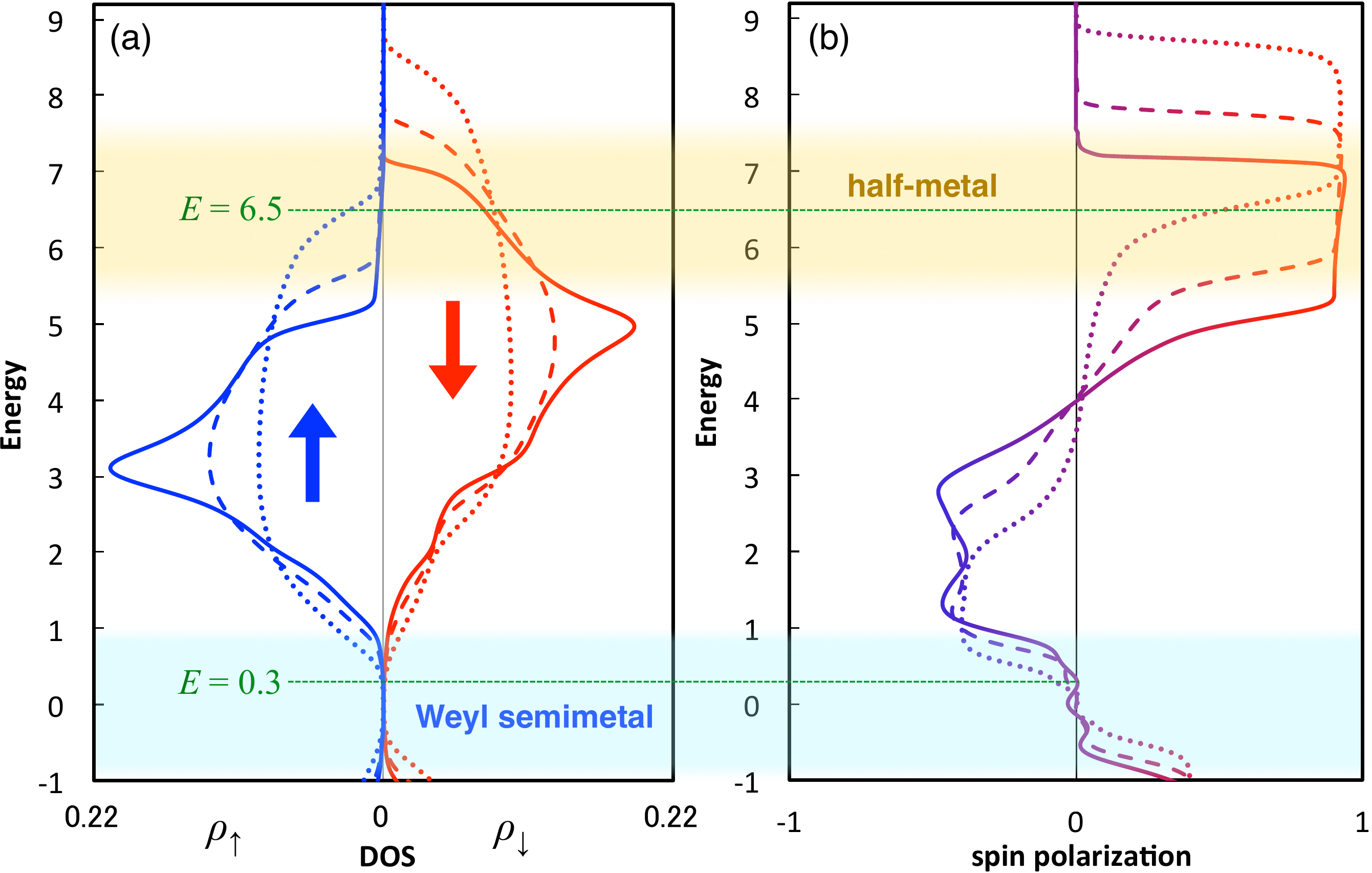}
 \vspace{-6mm}
\caption{
  (a) Spin-projected density of states $\rho_{\uparrow / \downarrow}$ and
 (b) spin-polarization as a function of $E$
 in a uniformly ($+z$) magnetized system
 for different disorder strengths of $W=1$ (solid lines), $W=3$ (dashed lines), and $W=5$ (dotted lines).
}
\label{fig:DOS}
\end{figure}

 Now we compare 
this magnetoresistance effect in WSMs 
with the conventional effect
in highly spin-polarized metals,
i.e., the magnetoresistance due to spin mistracking.
 The latter is also demonstrated in our model Hamiltonian Eq.~\eqn{H_real}
when the Fermi level is near the band edge.
 To observe this,
we calculate the 
spin-projected density of states, $\rho_{\uparrow}$ and $\rho_{\downarrow}$,
with uniform magnetization in the $z$ direction [see Fig.~\ref{fig:DOS}(a)], 
using the kernel polynomial method \cite{Weisse06}.
 Plotting 
the spin polarization 
${\rho_{\uparrow}- \rho_{\downarrow} \over \rho_{\uparrow}+ \rho_{\downarrow}}$
as in Fig.~\ref{fig:DOS}(b),
it becomes clear that
an almost perfectly polarized half-metal state is obtained
near the upper band edge ($E\simeq 7$).
 Therefore, just by changing the Fermi energy,
we can compare the WSM 
and the ideal half-metal 
within the same model.

\begin{figure}[!tbp]
 \centering
  \includegraphics[width=1\linewidth, bb =0 0 651 283]{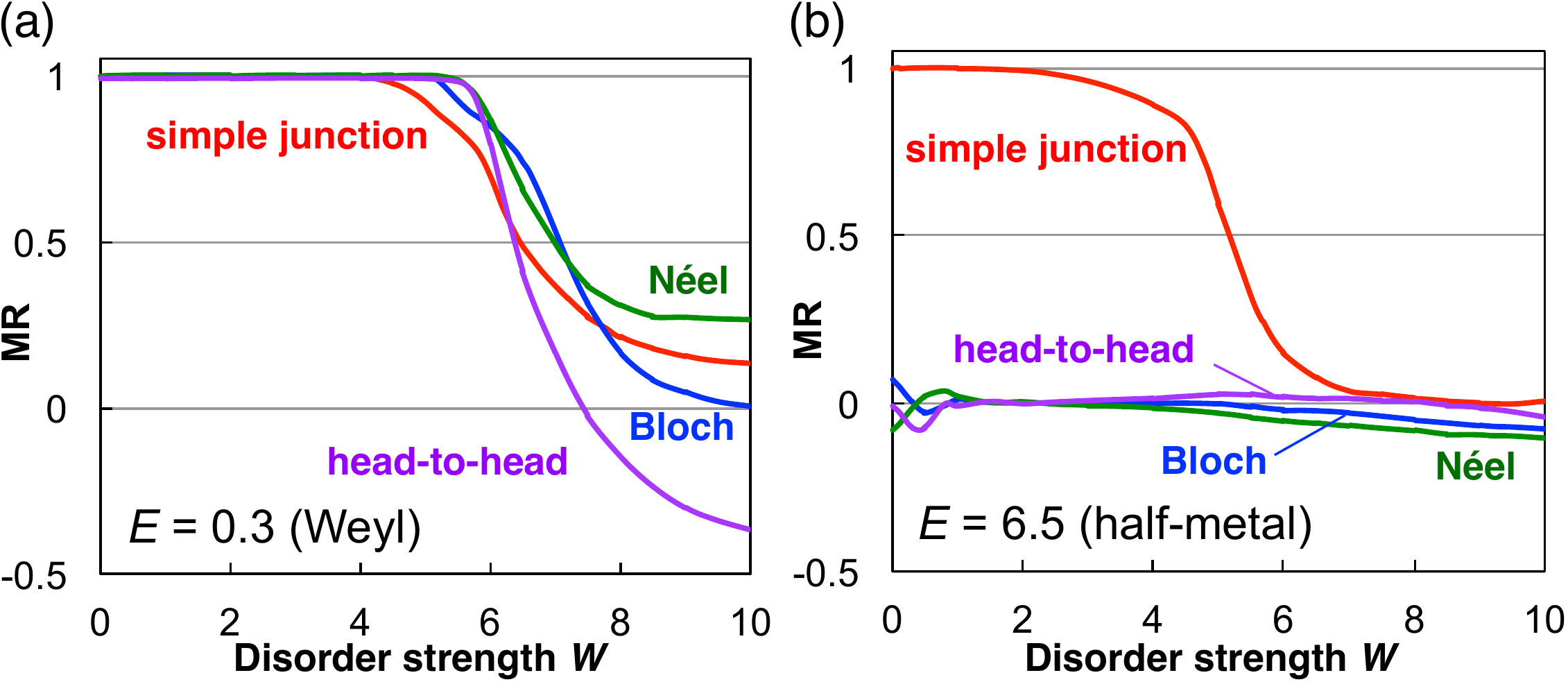}
 \vspace{-6mm}
\caption{
  MR as a function of disorder strength $W$,
 in (red) simple junction, (blue) Bloch, (green) N{\'e}el, and (purple) head-to-head domain walls,
 with $\theta\_{DW}=\pi$
 for (a) $E=0.3$, where Weyl cones appear, and 
 (b) $E=6.5$, where ideal half-metals arise.
  The size was set to $N=24$.
}
\label{fig:MRenergy}
\end{figure}

 Figure~\ref{fig:MRenergy} shows the MR (a) in WSMs ($E=0.3$) and (b) in ideal half-metals ($E=6.5$) with various types of domain walls.
 At weak disorder, the MR in a WSM is almost $100\%$ for any type of domain wall.
 In contrast, 
the MR in a half-metal with
a domain wall is significantly suppressed,
while that for 
an antiparallel junction is almost $100\%$
as that in WSMs.
 This is one of the most important results in this work.
 In conventional half-metals with sufficiently thick (more than four lattice sites in this case) domain walls, 
the spins of conduction electrons can track the direction of local magnetization along the domain walls,
and thus the domain-wall magnetoresistance becomes negligibly small.
 On the other hand, 
in the WSM phase, 
the Weyl fermions in the $+$~and~$-$ nodes behave independently, and the helicity is conserved
[see Fig.~\ref{fig:schematic}(c)],
as long as the inter-node scattering is negligibly weak
\cite{note3}.
 The domain walls do not induce inter-node scattering 
in principle,
and thus the helicity mismatch nature leads to the perfect magnetoresistance, ${\rm MR} = 1$,
even in the presence of the (Bloch or N{\'e}el type) domain walls.
(Note that in the head-to-head wall case, the magnetoresistance comes from the less-overlapping of the Fermi surfaces.)
 Another important point is the robustness against disorder.
 The MR in a half-metal gradually decreases 
as disorder increases,
while that in a WSM remains unity at weak disorder
and abruptly decays near the WSM/metal transition point.

\begin{figure}[tbp]
 \centering
  \includegraphics[width=1\linewidth, bb =0 0 669 295]{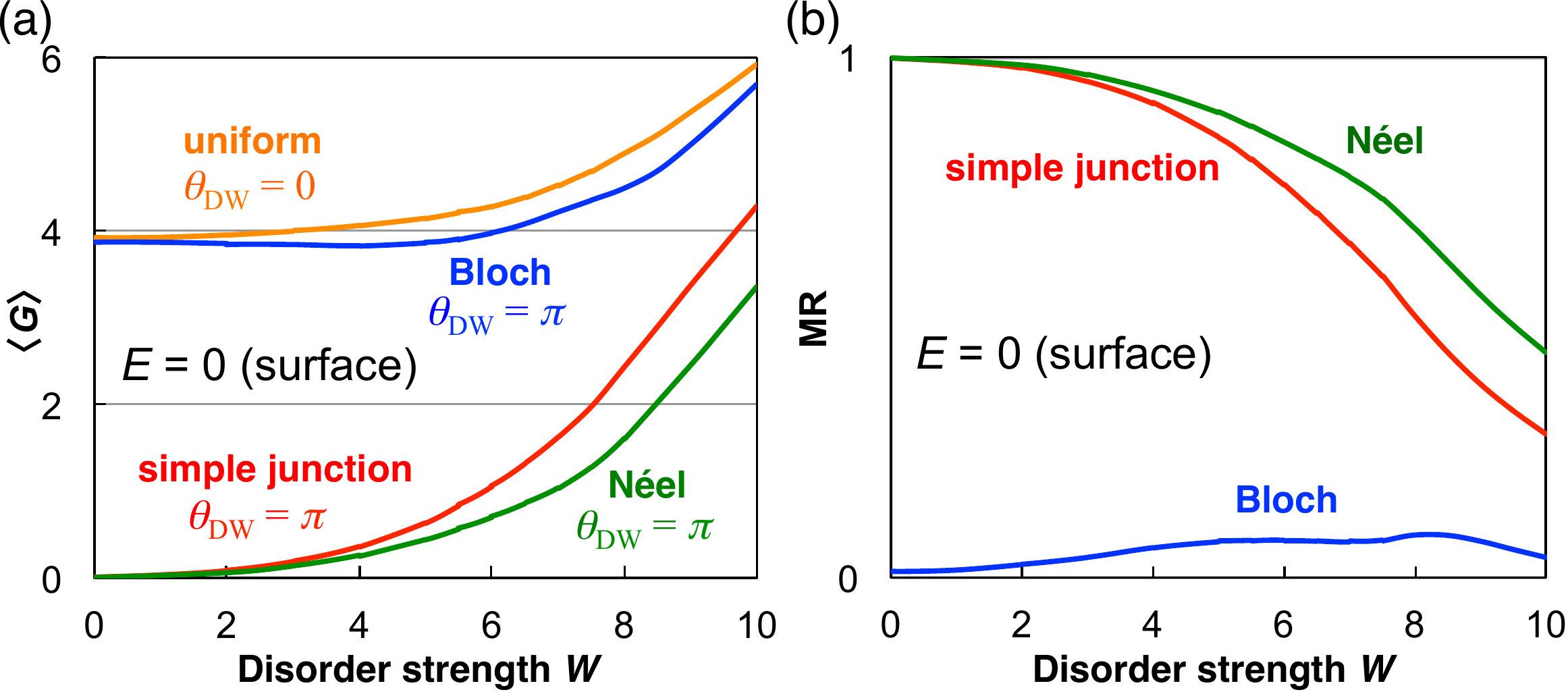}
 \vspace{-6mm}
\caption{
  (a) Averaged conductance $\braket{G}$ and (b) MR as functions of disorder strength $W$,
 for (orange) the uniform case, (red) antiparallel junction, (blue) Bloch wall, and (green) N{\'e}el wall
 with $N=24$ and Fermi energy $E=0$.
  Here the head-to-head wall is omitted 
 because the surface states do not carry current in the direction parallel to the magnetization.
}
\label{fig:G-Wsurf}
\end{figure}

\begin{figure}[tbp]
 \centering
  \includegraphics[width=60mm, bb =0 0 455 300]{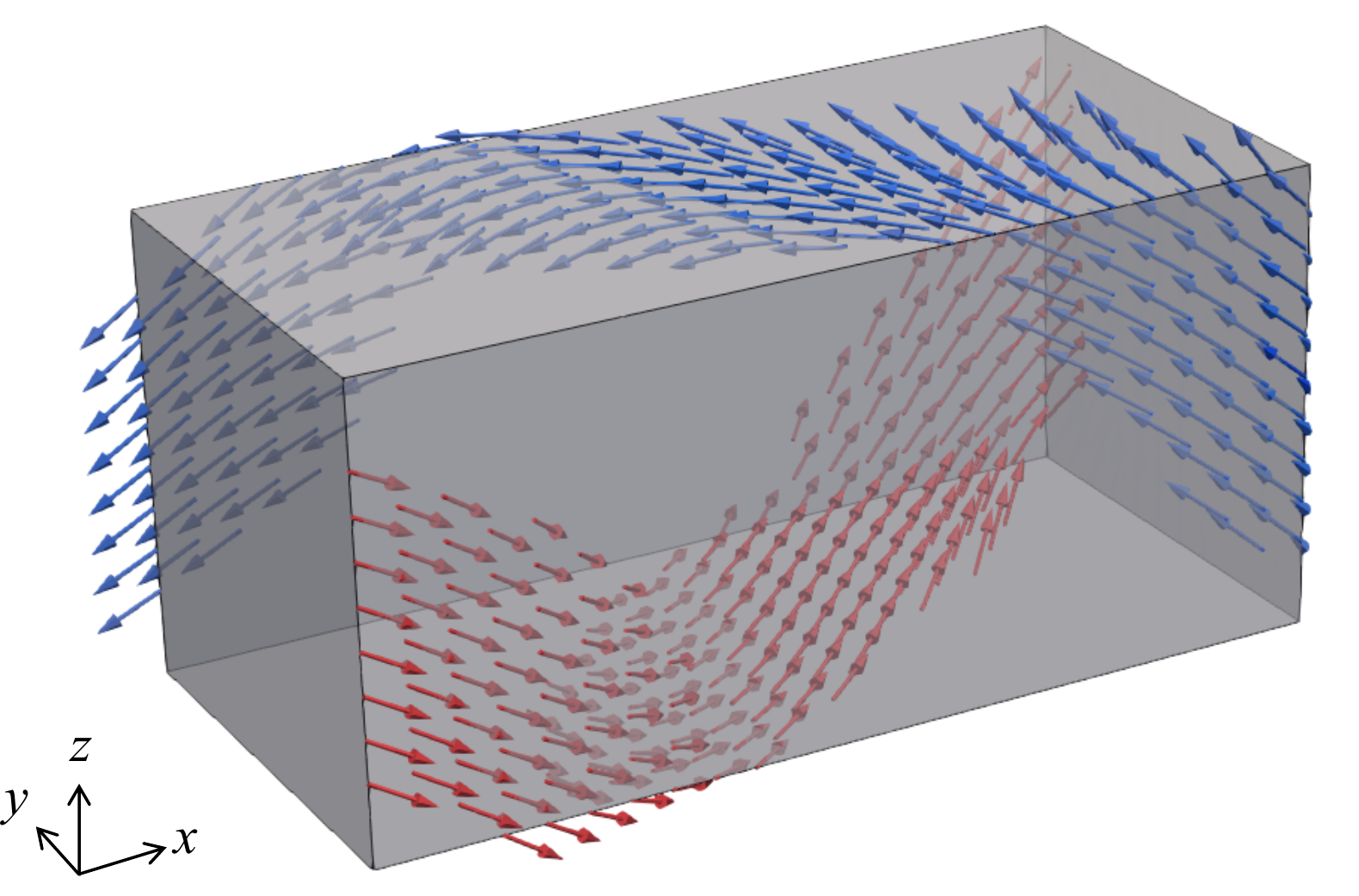}
 \vspace{-3mm}
\caption{
  Conducting state in a system with a Bloch wall in the clean limit.
  A pair of surface states appears along the spiral magnetization.
  Arrows indicate the direction of spin (${\hat{\bm s}}=\bra{\psi}\tau_0 {\bm \sigma}\ket{\psi}/\braket{\psi|\psi}$).
  For visibility, the length of the sample is doubled ($24\times 12\times 12$ sites), although in the main text we consider cubic samples.
}
\label{fig:spiral}
\end{figure}

 Before concluding, we discuss the contribution of the surface, i.e., Fermi arcs.
 When the system has surfaces,
almost gapless states appear on the surfaces parallel to the magnetization,
while the finite-size gap opens in the bulk.
 Therefore, the surface contribution becomes important for small samples.
 In our mesoscopic samples with surfaces, the low-energy ($E\simeq 0$) 
transport relies mostly on the surface states.
 On the surface, the spin
of the conducting state is locked with respect to the direction of magnetization and momentum
\cite{Lv15}.
 As a result, 
in a simple antiparallel junction, 
the current on the surface is completely reflected
and yields huge MR (see Fig.~\ref{fig:G-Wsurf}), as in the case of the bulk.
 However, with a Bloch domain wall, the conductance recovers and MR becomes small
because spiral surface states run through the system (Fig.~\ref{fig:spiral}).
 Therefore, 
the huge MR is achieved in the presence of surface states,
while only in the case of Bloch wall 
it may be suppressed as the contribution of surface states to transport increases.
 The surface magnetoresistance effect is not as robust against disorder as that for the bulk.
 This is considered to be due to the sensitivity of the surface spin-locking to disorder.

 We have studied the transport in disordered magnetic WSMs
and found a novel type of magnetoresistance effect that arises due to the helicity mismatch.
 Considering domain walls, we have shown that huge (almost $100\%$) MR is achieved, irrespective of the detail of the magnetization configuration.
 This is a particular feature of WSMs and
is in good contrast with conventional domain-wall magnetoresistance 
due to spin mistracking,
which is significantly suppressed 
by the introduction of a domain wall.
 We have studied the quantum transport in Weyl semimetals, and the $100\%$ MR will be obtained in mesoscopic systems at low temperatures.
 Although we have shown the data for small system sizes and the amplitude of the conductance in WSMs is of the order of $e^2/h$, 
the effect can be seen in larger system sizes \cite{sup1}.
 We note that the resistance effect occurs at the interface of two domains with antiparallel magnetizations, 
and the required condition is sufficiently long coherence length compared with the domain wall thickness (not with the device size). 
 While the resistance from less-overlapping is expected to be observed in ordinary ferromagnetic semiconductors \cite{Nguyen06}, it will be more prominent in WSMs due to the strong anisotropy of Fermi surfaces and robustness against disorder.
 The resistance from helicity-mismatch is a specific quantum transport phenomena for magnetic Weyl semimetals and will become important for spintronics devices.
 We also emphasize that the effect can be seen in a broad range of energy (i.e., it is not a singularity on the Weyl point $E = 0$), and it does not require a fine-tuning of chemical potential to observe. 
 We have also shown the robustness of the novel magnetoresistance effect against disorder.
 This robustness can be observed not only for non-magnetic potential disorder discussed here, but also for magnetic disorder
\cite{sup3}.
 These impurity and interface-roughness tolerant features should be advantageous for the manufacture of spintronics devices.
 Therefore, we propose that the magnetoresistance effect in ferromagnetic WSMs is more promising than the conventional GMR in ideal half-metals.

\begin{acknowledgements}
 This work was supported by 
KAKENHI Grants-in-Aid (Nos. JP15H05854, JP16J01981, and JP17K05485) from the Japan Society for the Promotion of Science (JSPS).
\end{acknowledgements}


\begin{thebibliography}{10}
\bibitem{note1}
 When introducing disorder, the renormalization of mass must be taken into account \cite{Groth09,Yamakage11,Kobayashi13}.
 In this paper, we tune the mass $m_0$ 
for a finite disorder strength $W$
so that the renormalized mass $\tilde m_0=0$, and
the Dirac semimetal line 
(i.e., the phase boundary of strong topological insulator/ordinary insulator)
in the phase diagram in Ref.~\onlinecite{Kobayashi13} is then traced.
\bibitem{note2}
Assuming the magnetization
$\bm M$ is uniform and oriented in the $\nu$ direction, a pair of Weyl nodes appears at $k_\nu=\pm k_0$,
\begin{equation*}
 k_0 \!=\!
       \cos^{-1}\!\[ \frac{m_\nu m_2\pm\! \sqrt{m_\nu^2 t^2 + (m_2^2 \!-\! t^2)(t^2 \!-\! M^2)}}{m_2^2 - t^2}\],
\label{k0}
\end{equation*}
\begin{equation*}
  m_\nu = m_0 + m_{2}\(3 - \sum_{\mu\ne \nu} \cos k_\mu\),  \quad k_\mu=0\ \textrm{or}\ \pi,
\label{eqn:m_bulk}
\end{equation*}
if $k_0$ has a real solution.
\bibitem{note3}
 In the continuum model, this results from the conservation of the chiral charge $n_{+}-n_{-}$, where $n_{\pm}$ is the number of Weyl electrons with $\pm$ helicity \cite{cxLiu13}.
\end{thebibliography}

\pagebreak
\thispagestyle{empty}
\widetext
\centering
\includegraphics[width=180mm, bb =40 0 565 772]{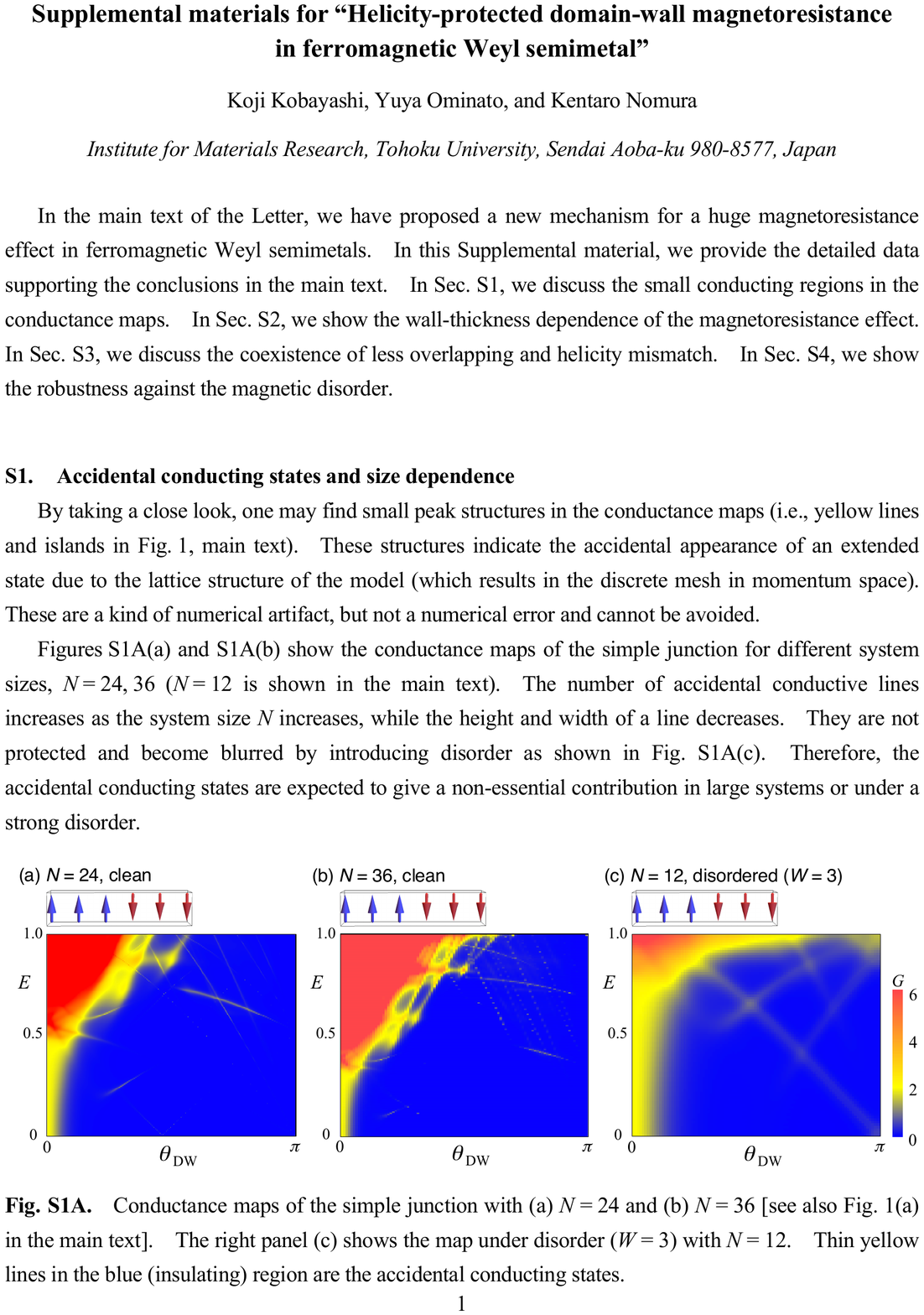}

\pagebreak
\thispagestyle{empty}
\widetext
\centering
\includegraphics[width=180mm, bb =40 0 565 772]{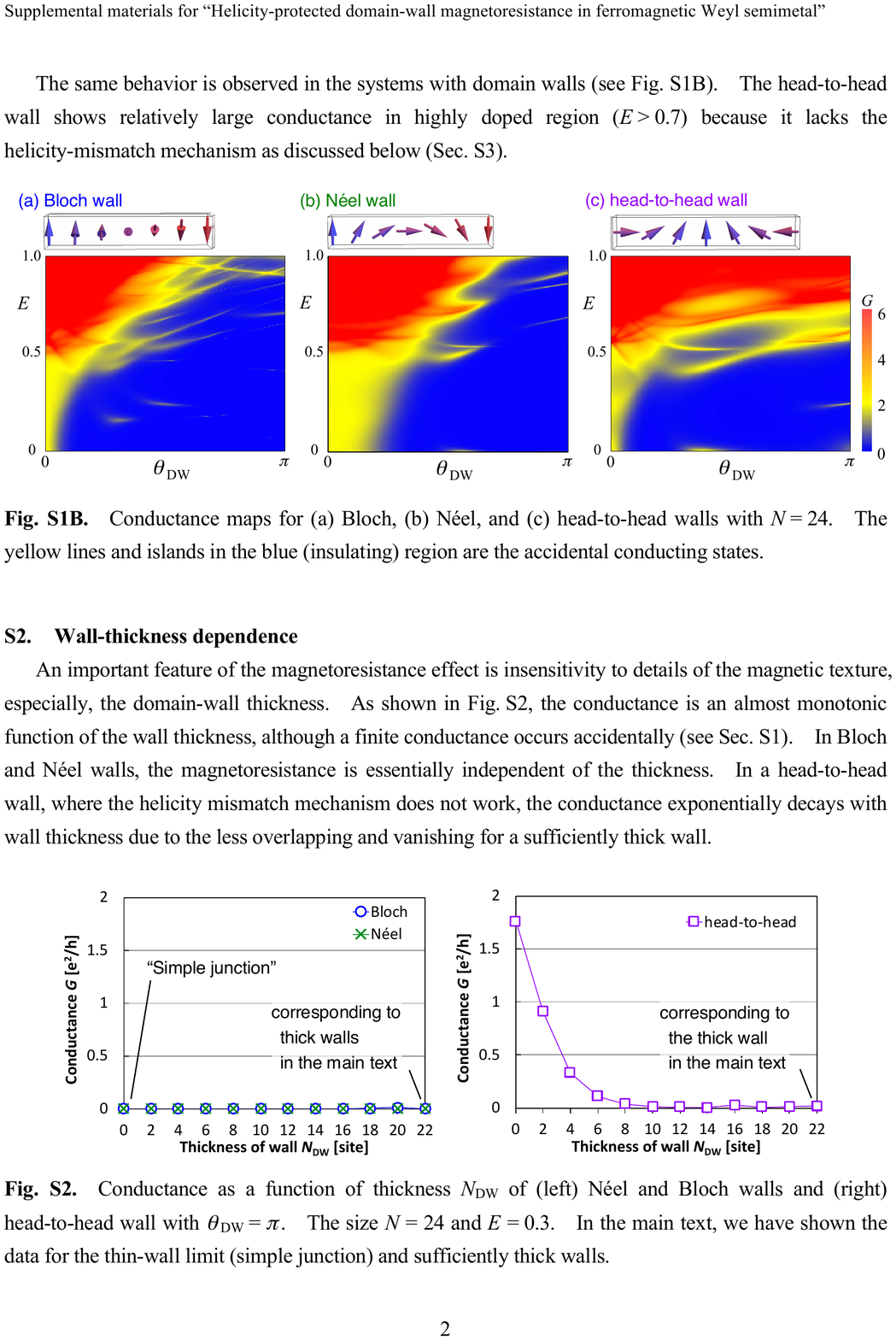}

\pagebreak
\thispagestyle{empty}
\widetext
\centering
\includegraphics[width=180mm, bb =40 0 565 772]{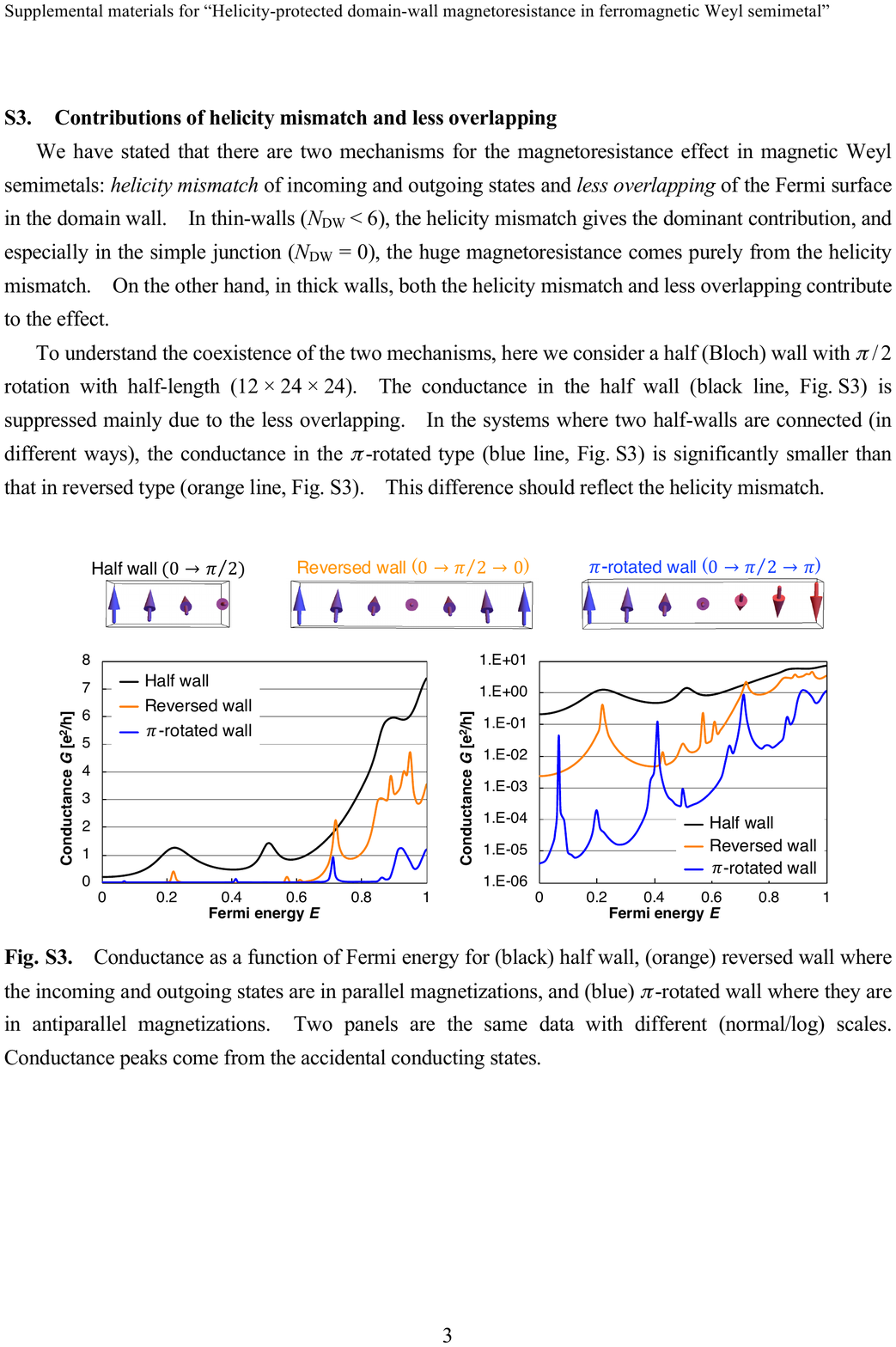}

\pagebreak
\thispagestyle{empty}
\widetext
\centering
\includegraphics[width=180mm, bb =40 0 565 772]{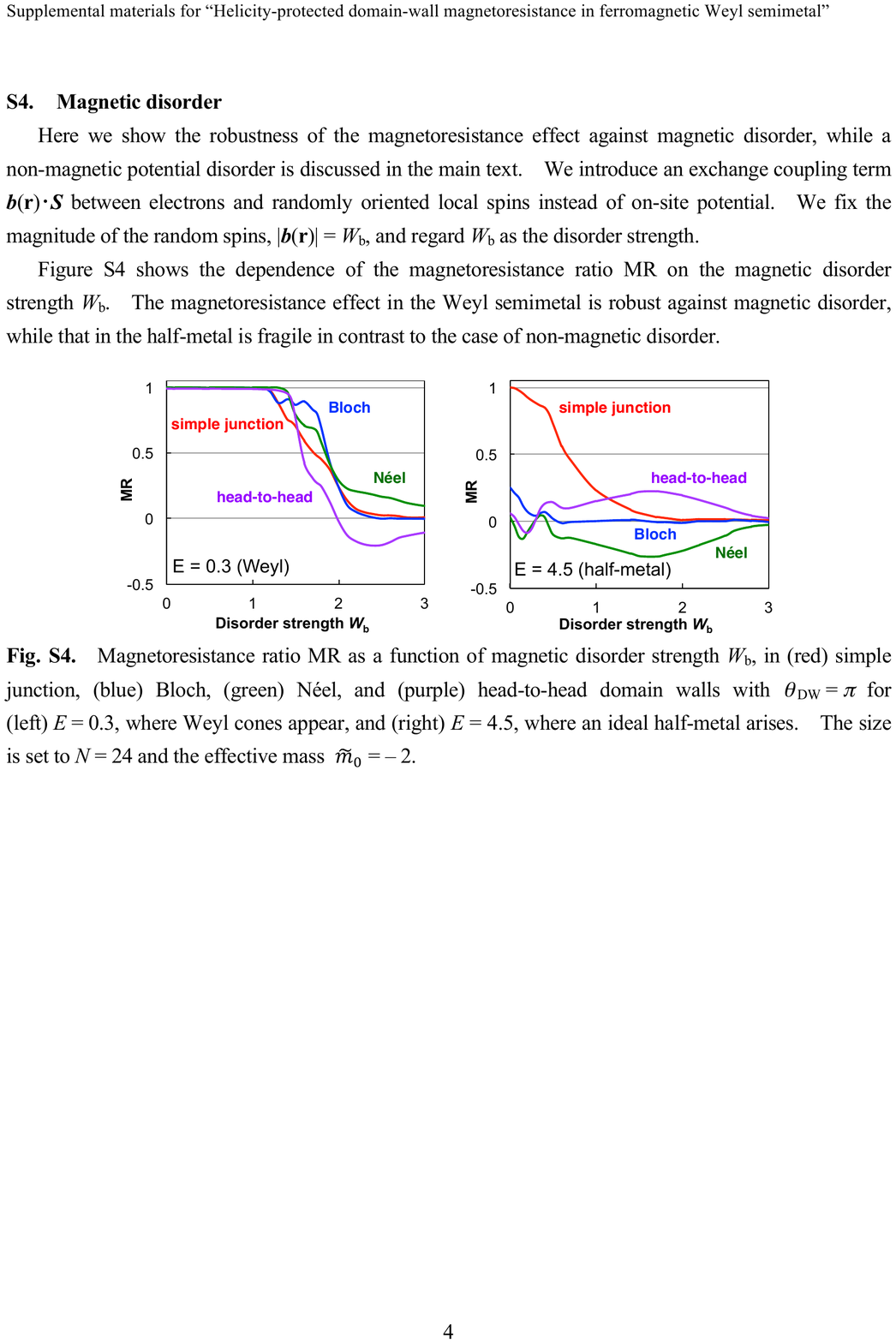}

\end{document}